\documentclass[12pt]{article}
\usepackage[margin=1in,footskip=0.25in]{geometry}
\usepackage{amsmath, amssymb, graphicx, epstopdf}
\usepackage{braket}

\numberwithin{equation}{section}
\newcommand\numberthis{\addtocounter{equation}{1}\tag{\theequation}}

\renewcommand{\bra}{\langle}
\renewcommand{\ket}{\rangle}

\title{Connecting phase transition theory with unsupervised learning}
\author{Kun Huang}
\date{\today}

\begin{document}
\maketitle

\begin{abstract}
Entropy and order parameter are two key concepts in phase transition theory. This paper proposes an unified method to both find order parameter and estimate entropy automatically with unsupervised learning. The contributions of this paper are threefold: First, it is shown that the cross-entropy loss of an optimum autoencoder could be used to estimate the physical entropy, which also explains why the critical temperature can be identified by the inflection point of the reconstruction loss. Second, a series of interpretable autoencoders are proposed which reproduce the ferromagnetic/antiferromagnetic (F/AF) order parameter in special cases. They provide us an intuitive prototype to understand the connection between unsupervised learning and phase transition theory. Third, we analyze spin glass phase with our method, the results suggest a ``distributed" order parameter to describe spin-glass ground state, which is a natural generalization of F/AF order parameter with respect to the autoencoder.
\end{abstract}
\section{Introduction}


Phase transition phenomenon can be more generally described as a process during which the complexity of the system undergoes a rapid change with some conditional parameters, such as temperature. Between the phases across which the transition happens, the one with less complexity is referred as ``ordered phase", the other referred as ``disordered phase".

As an example, consider a cooling process of an ferromagnetic/antiferromagnetic (F/AF) Ising lattice from above critical temperature $T_c$ to below $T_c$, the declining of ``complexity" from thermal disordered phase to ordered phase is closely related to the loss of ergodicity of the system in the phase space of microscopic states. In physics, the ``loss of ergodicity" is usually referred as ``symmetry breaking".

The ergodicity (or the``complexity") of a system can be formally described by its entropy, which is a functional of the ensemble distribution, $s=-\int{\rm d}x\rho(x)\ln\rho(x)$ (The integral runs over the phase space). Therefore, entropy is a key concept to understand phase transition, which exhibits a rapid change near $T_c$. (Though energy, or other thermodynamic quantities also exhibit rapid changes near $T_c$, these changes, in some sense, are just ``side effects", and may not cast light on the underlying mechanism.) However, it is difficult to evaluate entropy, except for some exactly solvable models. We also notice that \cite{Sauerwein:1995} proposed a Monte Carlo (MC) method to calculate the entropy for layer-wise Ising lattice with the help of transfer matrix, but MC is in general not suitable for estimating entropy, because entropy is not a direct function of the microscopic state. Thus, a more general method to estimate entropy is desirable.

Order parameter is another key concept in phase transition theory. Finding the order parameter is usually the primary task to describe a phase transition. Recently, \cite{LeiWang:2016} points out the order-parameter of F/AF Ising model is related to the low dimensional representation of the MC samples in a principal component analysis (PCA), and phase transition can be discovered by a clustering analysis. \cite{Wetzel:2017} uses various autoencoders to analyze F/AF order in Ising lattice, and propose that the critical temperature $T_c$ can be identified by the inflection point of the reconstruction loss.

In this paper, we propose an unified method to both find order parameter and estimate entropy automatically with unsupervised learning. Our contributions are the following:
First, we show that the cross-entropy loss of the optimum autoencoder is an estimator of the physical entropy, which also explains why $T_c$ can be identified by the inflection point of the reconstruction loss\cite{Wetzel:2017}. Second, we propose a simple autoencoder which can exactly reproduce the F/AF order parameter, equivalent to the principal component analysis method\cite{LeiWang:2016}, and provide us an intuitive prototype to understand the connection between unsupervised learning and phase transition theory.  Third, we analyze AF trianglar lattice, of which the ground state is a spin glass due to frustration. The results suggest a ``distributed" order parameter to describe spin glass, which is a natural generalization of F/AF order parameter with respect to the autoencoder.

\section{Numerical Results}

\begin{figure}
  \centering
  \includegraphics[width=\columnwidth]{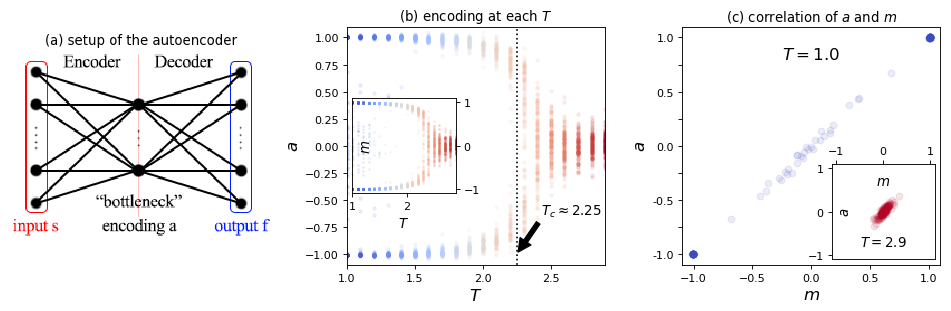}
  \caption{
  (a) The setup of the autoencoder $A_k$. The encoder consists of a single layer of linear neurons without biases, i.e, $a=W\cdot s$, where $s$ is the input vector which take values $\{1=\uparrow, -1=\downarrow\}$ at each of its $L=32\times32$ slots. $a$ is the encoding representation, of which the dimension $k$ is also referred as the ``bottleneck width", and the encoding weights $W$ is a $k\times L$ matrix. The sigmoid output of the decoder is denoted as $f$.
  (b) The encoding representation at different temperatures. Red dots correspond to the samples at higher temperatures, while blue dots at lower temperatures. The thickness of the color reflects the concentration of the samples. The dots exhibit a clear separation below $T_c\approx 2.25$. The inset graph shows the AF order-parameter $m$ of the samples, which gives the same behavior as the encoding representation. (c) A direct correlation between the encoding representation and the order-parameter. The blue dots gives the result for $T=1.0<T_c$, red dots of the inset for $T=2.9>T_c$. Panel (b) and (c) strongly indicate that the encoding representation is conceptually the same as the order parameter.}
  \label{fig1:encoded}
\end{figure}

As pointed out by \cite{LeiWang:2016}, the order-parameter of F/AF Ising lattice can be considered as the low dimensional representation of the MC samples in a principal component analysis (PCA), and it is well-known that PCA is equivalent to a linear autoencoder with a loss function of mean square error (MSE). Thus, it is natural to generalize PCA method to a more powerful autoencoder formulism, such as \cite{Wetzel:2017}. In this paper, we require the encoder to be linear, because we expect this encoder to be interpretable, and reproduce the PCA result in F/AF cases. Considering the fact that a multilayered linear encoder can always be collapsed into a single layer\cite{Hornik:1989}, thus the depth of the linear encoder is restricted to be 1. As shown in Figure~\ref{fig1:encoded}(a), the decoder is also chosen to have one layer to match the single-layered encoder. However, we replace the MSE loss-function in traditional linear-encoder with a cross-entropy loss-function, because we will show that cross-entropy loss of the optimum autoencoder is a natural estimator of the entropy of the system from which the data is sampled. The fitting capacity of the autoencoder could be increased by increasing the bottleneck width $k$, i.e, the dimension of the encoding representation. In the following, the autoencoder with a bottleneck width $k$ is referred as $A_k$.


As the first example, we use $A_1$ to analyse the Ising model $H=-J\sum\limits_{\bra ij\ket}s_i s_j$ on a square lattice with periodic boundary condition, where $s_i\in\{+1=\uparrow,-1=\downarrow\}$ denotes the spin on site $i$. $\bra ij\ket$ indicates a summation over nearest neighbors. Take coupling strength $J$ as the energy unit, such that $J=\pm1$ corresponds to the F/AF case respectively. This famous model is exactly solvable\cite{Onsager:1944}, and undergoes a phase transition near the critical temperature $T_c=2/\log(1+\sqrt{2})\approx2.27$.
To prepare feeding samples for the autoencoder, we take a $32\times32$ AF square lattice, thus the input size is $L=32\times32=1024$. We generate $100\times20$ independent spin configurations uniformly at temperatures $T=1.0,1.1,1.2,\cdots,2.9$ by MC algorithm\cite{Wolff:1989}, and feed them into the autoencoder $A_1$ as the training set, and generate another $100\times20$ independent samples as the validation set.

\begin{figure}
  \centering
  \includegraphics[width=\columnwidth]{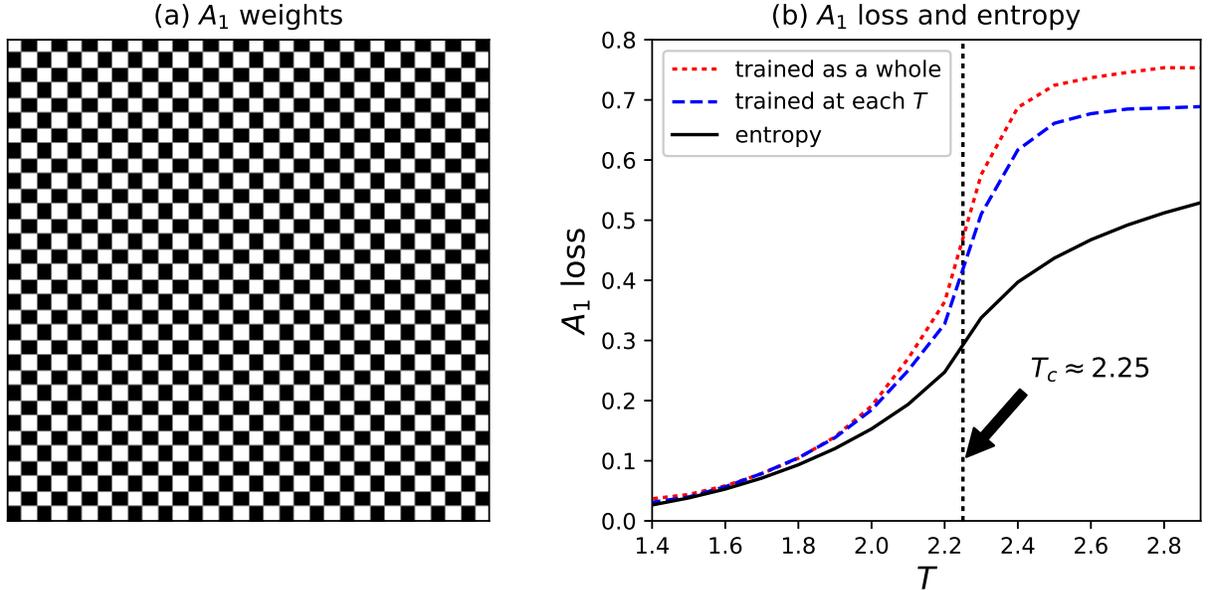}
  \caption{(a) The weights of $A_1$ for AF square lattice at $T=1.0$. The white squares are the sites where the values are positive, and black squares negative. The checkerboard pattern of the weights is consistent with the definition of AF order parameter. (b) The validation loss of $A_1$ for AF square lattice. The critical temperature $T_c$ can be identified by the inflection point of the losses. The red-dotted line gives the result when the samples at each temperatures are trained together. The blue-dashed line for training the samples at each temperature separately. The black-solid line gives the exact entropy of the AF square lattice.}
  \label{fig2:weight}
\end{figure}

Figure~\ref{fig1:encoded}(b) shows the encoding representation $a$ of the samples at each temperatures. The encoding representation $a$ exhibits a clear separation below $T_c\approx 2.25$. The inset of Figure~\ref{fig1:encoded}(b) gives the AF order parameter of the same samples, which is defined by $m={1\over L}\sum\limits_{p,q}{(-1)}^{p + q} s_{pq}$. $s_{pq}$ denotes the spin at $p$ row and $q$ column of the lattice. The behavior of $m$ is almost identical to the encoded representation $a$. Figure~\ref{fig1:encoded}(c) shows a direct correlation between $m$ and $a$ of the samples at $T=1.0$ and $T=2.9$.  Figure~\ref{fig2:weight}(a) gives the weights of $A_1$ encoder, which shows a checkerboard pattern, consistent with the definition of the AF order parameter. All these graphs clearly indicate that the encoding representation $a$ is conceptually the same as the order parameter $m$.  We also notice that it is possible to rescale the encoding weights of $A_1$ by multiplying an constant without changing the reconstruction loss\cite{Hornik:1989}. In physics, this means we are free to redefine the order-parameter by multiplying it with a constant.

Next, we look into the reconstruction lost $C$ (per site) of the validation set at each temperature, which is defined by,
\begin{equation}
C=-{1\over NL}\sum_{\sigma^{(n)} \in \text{samples}}\sum_i \left[\sigma^{(n)}_i\ln f_i^{(n)} + \left(1-\sigma^{(n)}_i\right)\ln\left(1-f_i^{(n)}\right) \right]
\end{equation}
where $\sigma^{(n)}_i\in\{0=\downarrow, 1=\uparrow\}$ is the ``one-hot" training target and $f^{(n)}_i$ is the sigmoid output of the decoder at site $i$. The superscript $(n)$ is the sample index. $L$ is the input size which equals to the number of lattice sites, and $N$ the samples number. The values of the mean validation loss at each temperature are given by the red dotted line in Figure~\ref{fig2:weight}(b). The critical temperature $T_c$ can be identified by the inflection point where the slope of the reconstruction loss is steepest. The black solid line in Figure~\ref{fig2:weight}(b) gives the values of the exact entropy\cite{Onsager:1944}. It is notable that the difference between the reconstruction loss of $A_1$ and the entropy is quite small for temperatures below $T_c$, though it grows larger for temperatures above $T_c$. In the Appendix, we show that for an ideal learning machine, the cross-entropy loss approximates to the entropy of the model from which the input data is sampled. Therefore, it is worth analyzing the deviation of the red line from the black line at high temperatures in Figure~\ref{fig2:weight}(b). There are two factors leading to the deviation. First, the training set of the red line consists of the samples of all temperatures. The samples at low temperatures are highly ordered, and resemble with each other much more than the samples at high temperatures. Due to mutual enhancement, the encoder captures more pattern information of the samples at low $T$s than high $T$s. Thus, the reconstruction loss at high $T$s would be smaller if we rule out the competition between the samples at low and high $T$s  by training the samples at each temperature separately. Besides, it seems also favorable in theory to train model at each $T$ separately, because the samples at different $T$ satisfy different distributions according to $\rho_T(s)=\exp(-\beta H)/Z$, where $Z$ is the partition function, and $\beta=1/T$. To verify this issue, we generate $2000$ uncorrelated samples at each $T$ for training (denoted as $D_T$), generate another $1000$ samples at each $T$ for validation (denoted as $V_T$).  and train $A_1$ with $D_T$ separately at each $T$. The resulting validation loss is given by the blue line in Figure~\ref{fig2:weight}(b). As expected, the loss is reduced at high temperatures, though still much larger than the entropy. This is because the construction loss also depends on the fitting capacity of the autoencoder itself.

As concluded in the Appendix, the more our learning machine fits the physical distribution, the more the cross-entropy loss of the machine agrees with the physical entropy. As for the autoencoders under consideration, $A_1$ is an ideal autoencoder in the limit $T\rightarrow 0$, because $D_0$ only consists of two kinds of ground-state configuration characterized by the two extreme values $\pm1$ of the 1d AF order-parameter. Thus an autoencoder with 1d encoding representation like $A_1$ is proper to describe samples $D_T$ when $T\rightarrow 0$. As $T$ increases, $A_1$ underfits $D_T$ more and more severely, due to lack of fitting capacity. We may use an autoencoder with a larger capacity than $A_1$, such as an $A_k$ with $k>1$, to avoid underfitting. However, if the fitting capacity is too large, we might meet another overfitting problem. Thus, we need to find an autoencoder $A^{(op)}$ with optimum capacity for each temperature, of which the reconstruction loss is expected to approximate the physical entropy.

In order to find $A^{(op)}$, we suppose, at each temperature, $A^{(op)}$ can be approximated by one of the autoencoders $\{A_k\}$, of which the fitting capacity can be tuned fine enough by varying the bottleneck width $k$. Our basic idea is that the best choice of the bottleneck width $k^{(op)}$ is characterized by a ``turning point" of the validation loss if we train the autoencoders for a sufficient long time.

\begin{figure}
  \centering
  \includegraphics[width=\columnwidth]{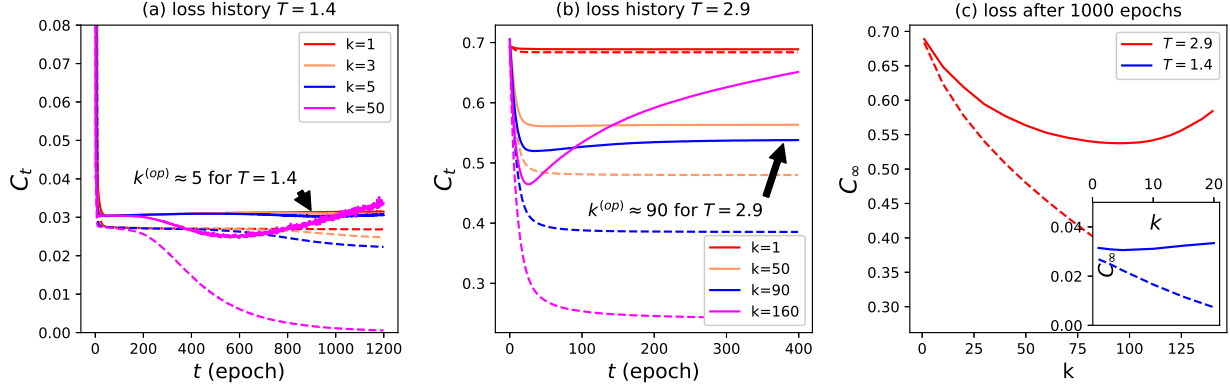}
  \caption{(a) The loss history of various $k$ at $T=1.4$ for AF square lattice. The solid/dashed line gives the loss of the validation/training set respectively. When $k\lesssim 5$,  the limiting value of the validation loss $C_\infty$ decreases consistently as $k$ increases. After the optimum value $k^{(op)}\approx 5$, $C_\infty$ climbs again if we increase $k$ furthermore.
  (b) The loss history of various $k$ at $T=2.9$ for AF square lattice. The optimum value $k^{(op)}\approx 90$ in this case. (c) $C_\infty$ as a function of fitting capacity, measured by the bottleneck width $k$. The solid/dashed line gives the limiting loss $C_\infty$ of the validation/training set respectively. As $k$ increases, The training loss decreases all the way to zero, while the validation loss has a ``turning point" $k^{(op)}$,  above which it climbs again, caused by overfitting the noise of training set. The red line is for $T=2.9$ with $k^{(op)}\approx 90$, and the blue line for $T=1.4$ with $k^{(op)}\approx 5$.}
  \label{fig3:history1}
\end{figure}

To be specific, we take the training process at $T=2.9$ as an example. As shown in Figure~\ref{fig3:history1}(b), After training for a sufficient long time $t$,  measured by the number of traing epochs, the validation loss $C_t$ approaches to a limiting value $C_\infty$. When $k$ is relative small, the fitting capacity of $A_k$ is limited. In this region, as we increase $k$, $C_\infty$ decreases consistently until a threshold value $k^{(op)}$ ($k^{(op)} \approx 90$ in our case). After $k^{(op)}$, $C_\infty$ climbs again if we increase $k$ furthermore.  For $k<k^{(op)}$, $C_\infty$ decreases as $k$ increases, which means the additional capacity is mostly used to capture the effective information in the training set. Thus, $A_k$ gets better at fitting the validation set, and we may refer the region $k<k^{(op)}$ as ``underfitting region". For $k>k^{(op)}$, $C_\infty$ increases again, which means the additional capacity begins to fit more of the noise in the training set, thus gets worse at fitting the validation set, and we may refer the region $k>k^{(op)}$ as ``overfitting region". As shown in Figure~\ref{fig3:history1}(c), we plot $C_\infty$ as a function of $k$ for each temperature. The optimum bottleneck width $k^{(op)}$ is then characterized by the turning point of the curve, which is the threshold value to separate the ``underfitting region" and the ``overfitting region".

\begin{figure}
  \centering
  \includegraphics[width=\columnwidth]{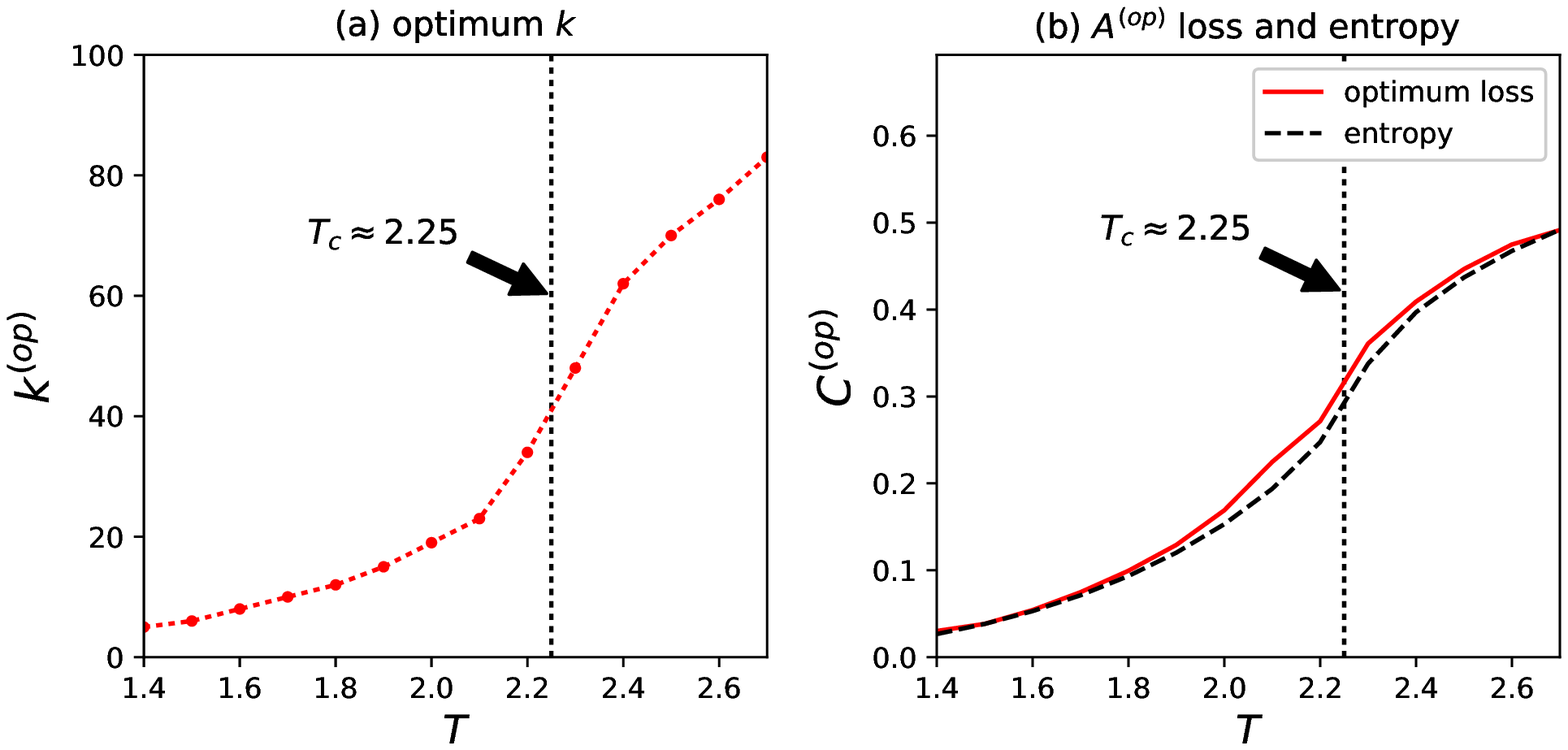}
  \caption{(a) The optimum dimension $k^{(op)}$ of the order parameter for AF square lattice at finite temperatures. $k^{(op)}$ approaches to $1$ as $T\rightarrow0$, which reproduces the traditional AF order parameter. For high temperatures, $k^{(op)}$ increases significantly. In particular, it climbs very fast near $T_c\approx 2.25$. (b) The validation loss of the optimum autoencoder $A^{(op)}$ for AF square lattice. The dashed line gives the exact value of entropy. The solid line gives the optimum validation loss, which matches the entropy very well.}
  \label{fig4:entropy2}
\end{figure}

Figure~\ref{fig4:entropy2}(a) gives the values of $k^{(op)}$ at each temperature for AF square lattice.  $k^{(op)}$ increases as $T$ increases, and it climbs very fast near $T_c$. As discussed above, the bottleneck width $k$ of the encoding representation can be considered as the dimension of the order parameter. In this sense, we may conclude that an order parameter with a dimension $k^{(op)}>1$ might be a better choice to describe AF Ising lattice at finite temperatures. Figure~\ref{fig4:entropy2}(b) gives the validation loss for $A^{(op)}$, which is represented by $A_{k^{(op)}}$ at each temperature. As expected, the optimum validation loss $C^{(op)}$ agrees with the entropy of the system quite well.

\begin{figure}
  \centering
  \includegraphics[width=\columnwidth]{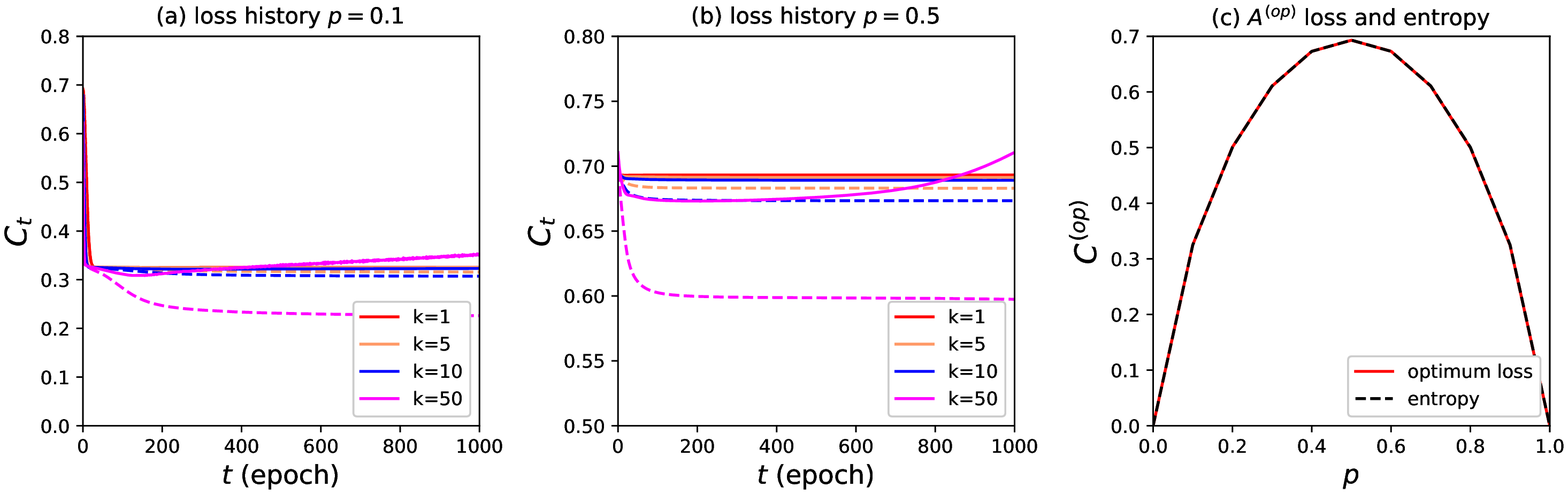}
  \caption{(a) The loss history of various $k$ at $p=0.1$ for the ``coin toss" model. The solid/dashed line gives the loss of the validation/training set respectively. The optimum dimension $k^{(op)}\approx 1$ in this case, and the validation loss has little change for small $k$ ($k\lesssim 10$). (b) The loss history of various $k$ at $p=0.5$ for the ``coin toss" model. Again, the validation loss has little change for $k\lesssim 10$, and we may choose $k^{(op)}\approx 1$ as well.(c) The validation loss of the optimum autoencoder $A^{(op)}$ for ``coin toss" model. The dashed line gives the exact value of ``coin toss" entropy. The solid line gives the optimum validation loss, which matches the entropy perfectly.}
  \label{fig5:entropy3}
\end{figure}

To further verify our method of finding $A^{(op)}$, we consider the paramagnetic limit when there is no coupling between the spins of square Ising lattice. We assume the probability of $s_i=\uparrow$ for every spin is $p$. Actually, this simple setup is equivalent to a ``coin toss" experiment, thus the exact entropy is given by $s=p\ln p + (1-p)\ln(1-p)$ as depicted by the dashed line in Figure~\ref{fig5:entropy3}(c). We take $1000$ training samples and $1000$ validation samples of the $32\times32$ lattice at each $p=0,0.1,0.2,\cdots,0.9,1.0$, and determine $k^{(op)}$ for each $p$ by finding the ``turning point" of the reconstruction loss. As shown in Figure~\ref{fig5:entropy3}(a) and (b), $k^{(op)}$ is small for every $p$ ($k^{(op)}\lesssim 10$), and the difference of the reconstruction loss is negligibly small for every $k<10$. The red line in Figure~\ref{fig5:entropy3}(a) gives the reconstruction loss $C^{(op)}$ as a function of the probability $p$, which perfectly agrees with the ``coin toss" entropy.

\begin{figure}
  \centering
  \includegraphics[width=\columnwidth]{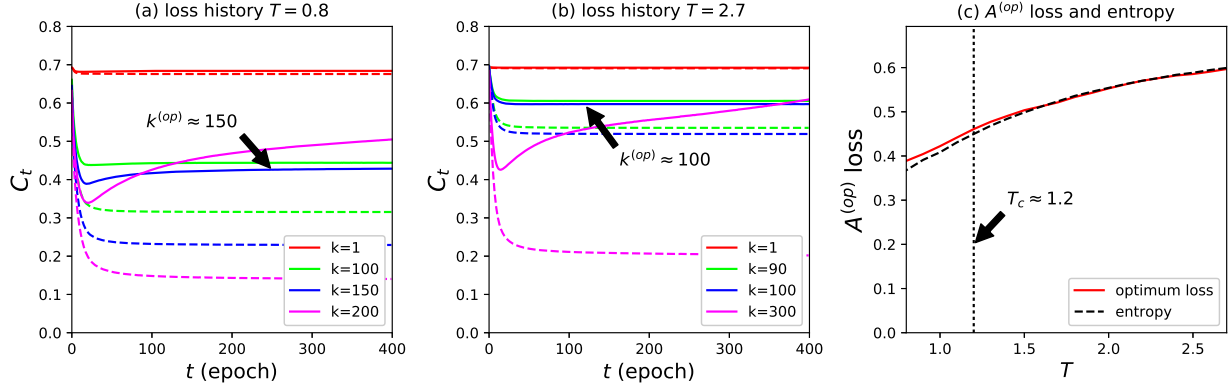}
  \caption{(a) The loss history of various $k$ at $T=0.8$ for AF triangular lattice, which has a spin-glass ground state caused by frustration. The solid/dashed line gives the loss of the validation/training set respectively. In this case, the optimum dimension $k^{(op)}\approx 150$ , which is much larger than the dimension of F/AF order parameter. (b) The loss history of various $k$ at $T=2.7$ for AF triangular lattice. The optimum dimension $k^{(op)}\approx 100$ in this case. (c) The validation loss of the optimum autoencoder $A^{(op)}$ for AF triangular lattice. The dashed line gives the exact value of entropy estimated by the transfer-matrix method\cite{Sauerwein:1995}. The solid line gives the optimum validation loss, which matches the entropy very well.}
  \label{fig6:entropy4}
\end{figure}

As the last example, we analyze the AF triangular lattice, of which the ground state is a spin glass due to frustration, and the critical temperature is $T_c\approx 1.2$\cite{Wannier:1950, Sauerwein:1995,TAKENGNY:2012}. We generate $4000$ training samples and $1000$ validation samples at each $T=0.8,0.9,\cdots,2.6,2.7$. As shown in Figure~\ref{fig6:entropy4}(a) and (b), $k^{(op)}$ is quite large even for $T<T_c$ ($k^{(op)}\sim 100$). This result implies that the configurations of spin-glass ground-state is highly degenerated, and its information should be represented by a high dimensional order parameter in a ``distributed" manner. This is quite different from the F/AF case of which the dimension of the ground-state order parameter is $k^{(op)}\sim 1$.  The reconstruction loss $C^{(op)}$ of AF triangular lattice is depicted by the red line in Figure~\ref{fig6:entropy4}(c), which is quite close to the physical entropy estimated from the transfer-matrix method \cite{Sauerwein:1995}, and the crital temperature for spin-glass transition $T_c\approx 1.2$ can be identified by a weak inflection point of $C^{(op)}$.

\section{Conclusion}

In brief, we develop an unsupervised-learning method to estimate entropy $S$ by searching for the optimum reconstruction loss $C^{(op)}$. Consequently, the critical temperature $T_c$ can also be identified by the inflection point of $C^{(op)}$. We emphasize that $C^{(op)}$ is an internal property of the physical system, irrelevant to the details of autoencoder itself. As the reconstruction loss, $C^{(op)}$ measures the complexity of the data samples in nature, while entropy $S$ also measures the complexity of the system. Therefore, $C^{(op)}$ and $S$ should have a definite relation, and we show their consistency both numerically and theoretically (See the Appendix). It is notable that, except for the data samples, our method of estimating entropy needs no priori knowledge of the Hamiltonian, which highlights the fact that entropy is a measure of complexity, and has no direct dependence of the Hamiltonian. At the same time, the definition of order parameter appears as a natural ``by-product" in our method, and the optimum dimension of the order parameter $k^{(op)}$ is closely related to the system complexity as well. In particular, a ``distributed" order parameter is proposed to describe spin-glass ground state, which is a natural generalization of F/AF order parameter in our method.

\section{Appendix}
In order to estimate the entropy of a physical system, such as Ising lattice, with machine learning, we suppose the distribution function $p(\sigma)$ is a marginal probability of $p(\sigma, h)$, where $\sigma$ denotes the microscopic state, i.e, the spin states for Ising lattice. $h$ denotes the parameters of the training model. (Different $h$ corresponds to different model candidates), then we have,
$$p(\sigma)\equiv\int {\rm d}h p(\sigma, h) = \int{\rm d}h p(\sigma|h)p(h)$$
Suppose we found an optimum model with parameters $h^*$ by some machine learning method, then we may take, $p(h)\approx \delta(h-h^*)$, where $\delta(\cdot)$ is the Dirac function, then, $p(\sigma)\approx p(\sigma|h^*)$. In other words, $\hat{p}(\sigma)\equiv p(\sigma|h^*) $ is an estimator of the physical distribution $p(\sigma)$. We choose some feed-forward network, and take a sigmoid output $f_i$ as the probability for $s_i=\uparrow$. In this setup, if $h^*$ is given, the spins are conditionally independent of each other, i.e,
\begin{align*}
\hat{p}(\sigma) &= \prod_i \hat{p}(\sigma_i|h^*)\numberthis\label{eq_3}\\
\hat{p}(\sigma_i|h^*) &\equiv \sigma_i f_i + (1-\sigma_i) (1-f_i)\numberthis\label{eq_3a}
\end{align*}
where $\sigma_i=1$ if the spin $s_i=\uparrow$, $0$ if $s_i=\downarrow$; $f_i$, denoting the sigmoid output at site $i$, depends on $h^*$. With $\hat{p}(\sigma)$, the estimator of the physical entropy per site is given by,

\begin{equation}\hat{S}\equiv -{1\over L} \sum_\sigma \hat{p}(\sigma)\ln \hat{p}(\sigma)\approx -{1\over NL}\sum_{\sigma^{(n)} \in \text{samples}} \ln \hat{p}\left(\sigma^{(n)}\right)\label{eq_4}\end{equation}

where $L$ is the sites number, $N$ the number of samples for training. Substitute equation [\ref{eq_3}] into [\ref{eq_4}], we obtain

\begin{align*}
\hat{S}&\approx -{1\over NL}\sum_{\sigma^{(n)} \in \text{samples}}\sum_i \ln \hat{p}\left(\left.\sigma_i^{(n)}\right|h^*\right)\\
&=-{1\over NL}\sum_{\sigma^{(n)} \in \text{samples}}\sum_i \left[\sigma^{(n)}_i\ln f_i^{(n)} + \left(1-\sigma^{(n)}_i\right)\ln\left(1-f_i^{(n)}\right) \right]\label{eq_5}
\end{align*}

This is exactly the definition of the sigmoid cross-entropy in machine learning. In other words, if the model $\hat{p}(\sigma|h^*)$, trained from data samples, is a reliable estimator of the ensemble distribution $p(\sigma)$,
then the cross-entropy outputted from the model is a proper estimator of the physical entropy.

\end{document}